\documentclass[twocolumn,preprintnumbers,amsmath,amssymb,prl,superscriptaddress]{revtex4}
\usepackage{graphicx}
\usepackage{dcolumn}
\usepackage{bm}
\usepackage{units}%
\usepackage{stmaryrd}%
\usepackage[breaklinks=true,pdfborder={0 0 0}]{hyperref}
\usepackage[usenames]{color}
\usepackage{tabularx}

\usepackage{color}
\usepackage{tikz}
\usepackage{multirow}

\begin{document}
\newlength{\LL} \LL 1\linewidth
\title{Quantum-Mechanical Relation between Atomic Dipole Polarizability \\ and the van der Waals Radius}
\author{Dmitry~V.~Fedorov}
\email[E-mail: ]{dmitry.fedorov@uni.lu}
\affiliation{Physics and Materials Science Research Unit, University of Luxembourg, L-1511 Luxembourg}
\author{Mainak~Sadhukhan}
\affiliation{Physics and Materials Science Research Unit, University of Luxembourg, L-1511 Luxembourg}
\author{Martin~St\"ohr}
\affiliation{Physics and Materials Science Research Unit, University of Luxembourg, L-1511 Luxembourg}
\author{Alexandre~Tkatchenko}
\affiliation{Physics and Materials Science Research Unit, University of Luxembourg, L-1511 Luxembourg}


\begin{abstract}
The atomic dipole polarizability, $\alpha$, and the van der Waals (vdW) radius, $R_{\rm vdW}$,
are two key quantities to describe vdW interactions between atoms in molecules and materials.
Until now, they have been determined independently and separately from each other.
Here, we derive the quantum-mechanical relation $R_{\rm vdW} = const. \times\alpha^{\nicefrac 17}$
which is markedly different from the common assumption $R_{\rm vdW} \propto \alpha^{\nicefrac 13}$
based on a classical picture of hard-sphere atoms. As shown for 72 chemical elements
between hydrogen and uranium, the obtained formula can be used as a unified
definition of the vdW radius solely in terms of the atomic polarizability. For vdW-bonded heteronuclear
dimers consisting of atoms $A$ and $B$, the combination rule $\alpha = (\alpha_A + \alpha_B)/2$
provides a remarkably accurate way to calculate their equilibrium interatomic distance.
The revealed scaling law allows to reduce the empiricism and improve the accuracy of
interatomic vdW potentials, at the same time suggesting the existence of a non-trivial relation 
between length and volume in quantum systems.
\end{abstract}

\maketitle
The idea to use a specific radius, describing a distance an atom maintains from other atoms in non-covalent interactions,
was introduced by Mack~\cite{Mack1932} and Magat~\cite{Magat1932}. Subsequently, it was employed by Kitaigorodskii
in his theory of close packing of molecules in crystals~\cite{Kitaigorodskii1955,Kitaigorodskii1971}.
This opened a wide area of applications related to the geometrical description of non-covalent bonds~\cite{Zefirov1989,Zefirov1995}.
The currently used concept of the vdW radius was formalized by Pauling~\cite{Pauling1960} and Bondi~\cite{Bondi1964}, who directly
related it to vdW interactions establishing its current name. They defined this radius as half of the distance between two atoms
of the same chemical element, at which Pauli exchange repulsion and London dispersion attraction forces exactly balance each other.
Since then, together with the atomic dipole polarizability, the vdW radius serves for atomistic description of vdW interactions in
many fields of science including molecular physics, crystal chemistry, nanotechnology, structural biology, and pharmacy.

The atomic dipole polarizability, a quantity related to the strength of the dispersion interaction, can be accurately determined
from both experiment and theory to an accuracy of a few percent for most elements in the periodic
table~\cite{Chu2004,Tkatchenko2009,Tkatchenko2012,Gobre2016,Gould2016,Hermann2017}.
In contrast, the determination of the atomic vdW radius is unambiguous for noble gases only, for which the vdW radius is defined
as half of the equilibrium distance in the corresponding vdW-bonded homonuclear dimer~\cite{Pauling1960,Bondi1964}.
For other chemical elements, the definition of vdW radius requires the consideration of molecular systems where the corresponding
atom exhibits a closed-shell behavior
similar to noble gases, in order to distinguish the vdW bonding from other interactions~\cite{Zefirov1989,Zefirov1995}.
Hence, a robust determination of vdW radii for most elements in the periodic table requires a painstaking analysis of a substantial
amount of experimental structural data~\cite{Batsanov2001}.

Consequently, from an experimental point of view, the vdW radius can only be considered as a statistical
quantity and available databases provide just \emph{recommended} values. Among them, the one proposed in 1964
by Bondi~\cite{Bondi1964} has been extensively used. However, it is based on a restricted amount of experimental
information available at that time. With the improvement of experimental techniques and increase of available data,
it became possible to derive more precise databases. A comprehensive analysis was performed by
Batsanov~\cite{Batsanov2001}. He provided a table of accurate atomic vdW radii for 65 chemical
elements serving here as a benchmark reference~\cite{Comment_1}. For noble
gases, missing in Ref.~\cite{Batsanov2001}, the vdW radii of Bondi~\cite{Bondi1964} are taken in our analysis~\cite{Comment_2}.
As a reference dataset for the atomic dipole polarizability, we use Table~A.1 of Ref.~\cite{Gobre2016}.
They are obtained with time-dependent density-functional theory and linear-response coupled-cluster calculations
providing an accuracy of a few percent, which is comparable to the variation among different sets of experimental
and theoretical results~\cite{Gould2016}.

The commonly used relation between the atomic dipole polarizability and the vdW radius is based on
a classical approach, wherein an atom is described as a positive point charge $q$ compensated by a uniform
electron density $(-3 q)/(4 \pi R_{\rm a}^3)$ within a hard sphere. Its radius $R_{\rm a}$ is identical to
the classical vdW radius. With an applied electric field $\mathcal{E}_{\rm ext}$, the point charge undergoes
a displacement $d$ with respect to the center of the sphere. From the force balance,
$q \mathcal{E}_{\rm ext} - q^2 d / R_{\rm a}^3 = 0$, and the definition of the dipole polarizability
via the induced dipole moment, $q d = \alpha \mathcal{E}_{\rm ext}$, it follows that
\begin{align}
\begin{array}{ll}
R_{\rm a} = \alpha^{\nicefrac 13}\ .
\end{array}
\label{eq.:RvdW_class}
\end{align}
This scaling law is widely used in literature relating the vdW radius to the polarizability.

In this Letter, we show that the quantum-mechanical (QM) relation between the two quantities is markedly
different from the classical formula. This result is obtained from the force balance between
vdW attraction and exchange-repulsion interactions considered within a simplified, yet realistic, QM model.
Our finding is supported by a detailed analysis of robust data for atomic polarizabilities and vdW
radii of 72 chemical elements.

\begin{table}[t!]
  \caption{For noble gases, the proportionality function of the QDO model given by
           Eq.~(\ref{eq.:Prop_func}) is shown versus its counterpart of real atoms.
           The results are obtained with $\mu\omega$ from Ref.~\cite{Jones2013,Comment_3}
           and the reference vdW radii~\cite{Bondi1964,Runeberg1998} and polarizabilities~\cite{Gobre2016}.
           All values are given in atomic units.}
\begin{ruledtabular}
\begin{tabular}{c||c|c|c|c|c}
Species & $\mu\omega$ & $R_{\rm vdW}^{\rm ref}$ & $\alpha^{\rm ref}$ & $C (\mu\omega, R_{\rm vdW}^{\rm ref})$ & $R_{\rm vdW}^{\rm ref} / (\alpha^{\rm ref})^{\nicefrac 17}$ \\
\hline
He   &   0.5178   &   2.65   & ~~1.38   &   ~2.33~   &   2.53  \\
Ne   &   0.4526   &   2.91   & ~~2.67   &   ~2.56~   &   2.53  \\
Ar   &   0.2196   &   3.55   & ~11.10   &   ~2.33~   &   2.52  \\
Kr   &   0.1778   &   3.82   & ~16.80   &   ~2.35~   &   2.55  \\
Xe   &   0.1309   &   4.08   & ~27.30   &   ~2.28~   &   2.54  \\
Rn   &   0.1092   &   4.23   & ~33.54   &   ~2.25~   &   2.56  \\
\end{tabular}
\end{ruledtabular}
\end{table}

Many properties of real atoms can be captured by physical models based on Gaussian wave functions~\cite{Bloch1997}.
Among them, the quantum Drude oscillator (QDO) model~\cite{Wang2001,Sommerfeld2005,Jones2013} serves as an insightful, efficient, and
accurate approach~\cite{Tkatchenko2012,Reilly2015,Tkatchenko2015,Sadhukhan2016,Gobre2016,Sadhukhan2017,Hermann2017}
for the description of the dispersion interaction. It provides
the dipole polarizability $\alpha \equiv \alpha_1 = q^2/\mu \omega^2$ expressed in terms of the three
parameters~\cite{Jones2013}: the charge $q$, the mass $\mu$, and the characteristic frequency $\omega$
modeling the response of valence electrons. The scaling laws obtained for dispersion coefficients
within the QDO model are applicable to accurately describe attractive interactions
between atoms and molecules~\cite{Tkatchenko2009,Tkatchenko2012,Jones2013,Gobre2016,Hermann2017}.
Here, we introduce the exchange--repulsion into this model to uncover a QM relation
between the polarizability and vdW radius. Motivated by the work of Pauling~\cite{Pauling1960}
and Bondi~\cite{Bondi1964}, we determine the latter from the condition
of the balance between exchange--repulsion and dispersion--attraction forces.
The modern theory of interatomic interactions~\cite{Stone-book} suggests that the equilibrium
binding between two atoms (including noble gases)
results from a complex interplay of many interactions. Among them, exchange--repulsion,
electrostatics, polarization, and dipolar as well as higher-order vdW dispersion interactions
are of importance.
However, it is also known that the Tang-Toennies model~\cite{Tang1995,Tang1998,Tang2003},
which consists purely of a dispersion attraction and an exchange repulsion, reproduces binding
energy curves of closed-shell dimers with remarkable accuracy. To express the vdW radius in terms
of the dipole polarizability, our initial model presented here treats the repulsive and attractive
forces by employing a dipole approximation for the Coulomb potential. Such an approximation turns
out to be reasonable to correctly describe the equilibrium distance for homonuclear closed-shell
dimers via the condition of vanishing interatomic force. Our dipolar QM model can also be generalized
to higher multipoles, as demonstrated by the excellent correlation between higher-order atomic
polarizabilities and the vdW radius (see Eq.~(\ref{eq.:Relation_general})).

\begin{figure}[t!]
\includegraphics[width=0.8\LL]{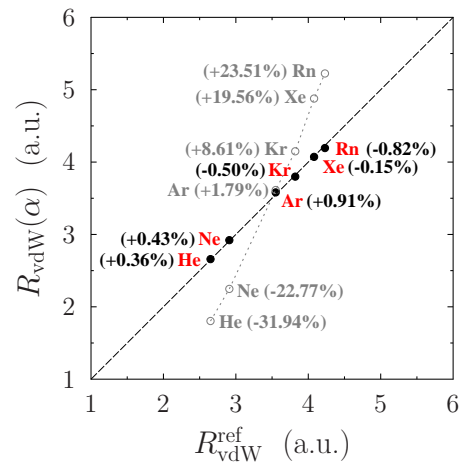}
\caption{(Color online)
The van der Waals radius obtained for noble gases by Eq.~(\ref{eq.:Relation_2}) is presented
in comparison to its reference~\cite{Batsanov2001} counterpart (black full dots). In addition,
the results obtained by the fit of the classical scaling law to the reference data, 
$R_{\rm vdW} (\alpha) = 1.62\,\alpha^{\nicefrac 13}$, are shown (grey circles).
The relative errors calculated with Eq.~(\ref{eq.:RE}) are in parentheses.}
\label{img:RvdW_NG}
\end{figure}

A coarse-grained QDO represents response properties of all valence electrons in an atom as those of a single oscillator~\cite{Jones2013}.
As a result, usual prescriptions to derive the Pauli exchange repulsion from the interaction of each electron pair~\cite{Tang1998}
are not straightforward within this model. However, two QDOs with the same parameters are indistinguishable. In addition, their spin-less
structure~\cite{Jones2013} is well suited to describe closed valence shells of atoms, which interact solely via the vdW forces. Considering
two identical QDOs as bosons, we construct the total wavefunction as a permanent and introduce the exchange interaction following
the Heitler-London approach~\cite{Heitler1927}, where it is expressed in terms of the Coulomb and exchange integrals.

\begin{figure*}[t!]
\includegraphics[width=0.8\LL]{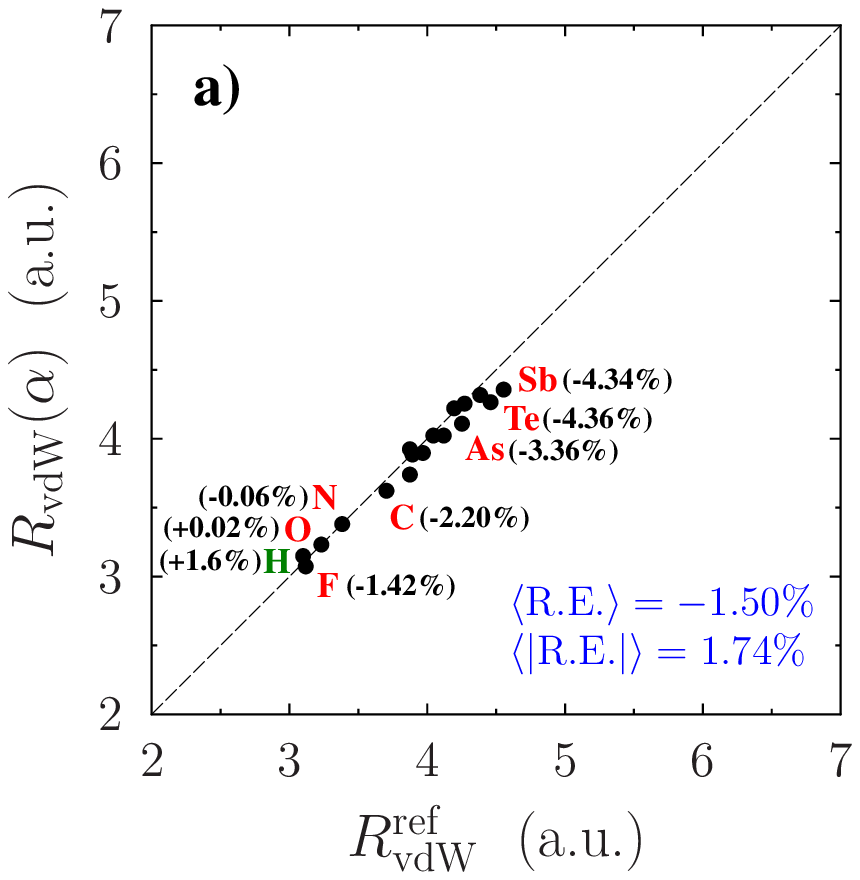}\qquad\qquad
\includegraphics[width=0.8\LL]{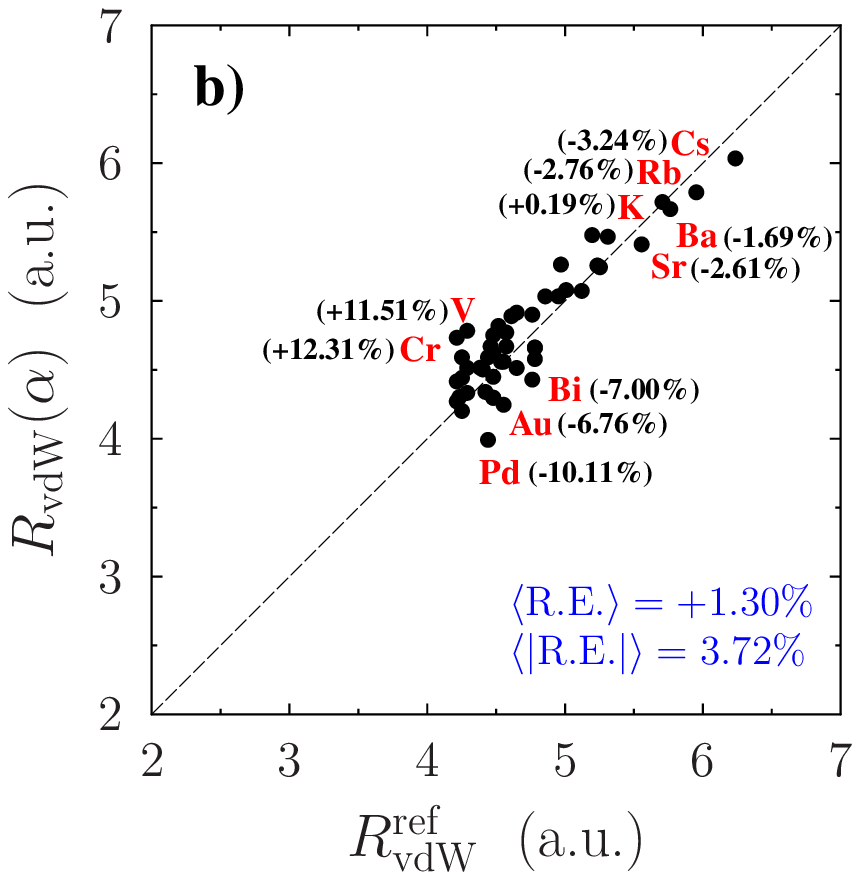}
\caption{(Color online)
The vdW radius obtained by Eq.~(\ref{eq.:Relation_2}) using the reference data for
the polarizability~\cite{Gobre2016} is shown separately for a) nonmetals/metalloids and b) metals in comparison
to its reference counterpart~\cite{Batsanov2001}. Here, $\langle\text{R.E.}\rangle$ and $\langle|\text{R.E.}|\rangle$
represent the mean of the relative error and its magnitude, respectively, calculated with Eq.~(\ref{eq.:RE}) for
the database of Batsanov~\cite{Comment_1}. In addition, we show the comparison between
$R_{\rm vdW} (\alpha)$ and $R_{\rm vdW}^{\rm ref}$ for H with the latter one taken from Ref.~\cite{Tkatchenko2009}.}
\label{img:RvdW_vs_RvdW}
\end{figure*}

Let us consider a homonuclear dimer consisting of two atoms separated by the distance $R$.
As shown in the Supplemental Material~\cite{Supplementary}, the dipole approximation for the Coulomb interaction
provides the exchange integral in the simple form
\begin{align}
\begin{array}{ll}
J_{\rm ex} = \frac{q^2 S}{2R} = \frac{q^2}{2R}~e^{-\frac{\mu\omega}{2\hbar} R^2}\ ,
\end{array}
\label{eq.:J_ex}
\end{align}
whereas the corresponding Coulomb integral vanishes. At the equilibrium distance,
$R = 2 R_{\rm vdW}$, of homonuclear dimers consisting of the species of Table~I, the overlap integral,
$S$ in Eq.~(\ref{eq.:J_ex}), is less than $0.02$~\cite{Supplementary}. In the first-order
approximation with respect to $S$, the exchange energy for the symmetric state, related to the bosonic nature
of the closed shells, is given by $J_{\rm ex}$~\cite{Supplementary}. As follows from Table~I,
for $R = 2 R_{\rm vdW}$, the condition $\frac{\mu\omega}{\hbar} \gg \frac{1}{R^2}$ is fulfilled.
Then, the corresponding force, $F_{\rm ex} = - \nabla_R J_{\rm ex}$, can be obtained as~\cite{Supplementary}
\begin{align}
\begin{array}{ll}
F_{\rm ex} \approx \frac{q^2}{2} \frac{\mu\omega}{\hbar}~e^{-\frac{\mu\omega}{2\hbar} R^2}
= \frac{\alpha \hbar\omega}{2} \left(\frac{\mu\omega}{\hbar}\right)^2~e^{-\frac{\mu\omega}{2\hbar} R^2}\ .
\end{array}
\end{align}

The attractive dipole-dipole dispersion interaction and the related force are known within the QDO model
as~\cite{Jones2013}
\begin{align}
\begin{array}{ll}
E_{\rm disp} = - \frac{3}{4} \frac{\alpha^2\hbar\omega}{R^6}\ \ \ \ \text{and}
\ \ \ \ F_{\rm disp} = - \frac{9}{2} \frac{\alpha^2\hbar\omega}{R^7}\ ,
\end{array}
\label{eq.:FvdW}
\end{align}
respectively. From $F_{\rm ex} + F_{\rm disp} = 0$, we get the relation
\begin{align}
\begin{array}{ll}
R_{\rm vdW} = C (\mu\omega, R_{\rm vdW})~\alpha^{\nicefrac 17}\ .
\end{array}
\label{eq.:Relation_1}
\end{align}
Here, the proportionality function~\cite{Comment_4}
\begin{align}
\begin{array}{ll}
C (\mu\omega, R_{\rm vdW}) = \frac{1}{2} \left(\frac{3\hbar}{\mu\omega}\right)^{\nicefrac 27}
\exp\left({\frac{2\mu\omega R_{\rm vdW}^2}{7\hbar}}\right)
\end{array}
\label{eq.:Prop_func}
\end{align}
depends on both $\mu\omega$ and $R_{\rm vdW}$. However, as shown by Table~I, the deviations of
$C (\mu\omega, R_{\rm vdW})$ from its mean value of 2.35 are within 9\% among different species.
This is in contrast to the strong variation of the model parameters by themselves.
Moreover, the actual ratio $R_{\rm vdW} / (\alpha)^{\nicefrac 17}$
is practically constant for all noble-gas atoms, according to the last column in Table~I.
By fitting the scaling law $R_{\rm vdW} \propto (\alpha)^{\nicefrac 17}$ to the reference data
for noble gases~\cite{Bondi1964,Runeberg1998}, we obtain a remarkable relation
\begin{align}
\begin{array}{ll}
R_{\rm vdW} (\alpha) = 2.54~\alpha^{\nicefrac 17}~\text{a.u.}
\end{array}
\label{eq.:Relation_2}
\end{align}
which is the central result of our work~\cite{Comment_5}.

The function $C (\mu\omega, R_{\rm vdW})$ corresponds to a universal scaling law between the atomic volume
and the electron density at $R_{\rm vdW}$~\cite{Comment_4}. Its deviations from 2.54 can be attributed to
the model simplifications related to the coarse-grained description of valence electrons by Gaussian wave functions.

Figure~\ref{img:RvdW_NG} shows that Eq.~\eqref{eq.:Relation_2} yields a relative error
\begin{align}
\begin{array}{ll}
\text{R.E.} = \left[ (R_{\rm vdW} (\alpha) - R_{\rm vdW}^{\rm ref})/R_{\rm vdW}^{\rm ref} \right] \times 100\%
\end{array}
\label{eq.:RE}
\end{align}
of less than 1\% for all noble gas atoms. In contrast,
the fit of the classical scaling law of Eq.~(\ref{eq.:RvdW_class}) to the reference data is clearly unreasonable.
The power law of Eq.~(\ref{eq.:Relation_2}) is also supported by our extended statistical analysis
performed for the noble gases by assuming different possible power laws~\cite{Supplementary}. Among them, the one
of Eq.~(\ref{eq.:Relation_2}) is identified as the actual scaling law with the coefficient of variation of
less than 1\% as well as the one with the minimal standard deviation.

Let us now assess the validity of the relation given by Eq.~(\ref{eq.:Relation_2}) for atoms of other chemical
elements. To this end, we use the equilibrium vdW radii of Batsanov~\cite{Batsanov2001} as the reference~\cite{Comment_1}.
For hydrogen, we take the value of the vdW radius from Ref.~\cite{Tkatchenko2009}.
The results of our analysis are illustrated in Fig.~\ref{img:RvdW_vs_RvdW}
separately for nonmetals/metalloids (16 elements of Ref.~\cite{Batsanov2001} + H) and metals (49 elements).
A detailed information is provided in the Supplemental Material~\cite{Supplementary}.
We observe an excellent correlation between $R_{\rm vdW} (\alpha)$ and its reference counterpart
for a wide range of input data: $1.38 \leqslant \alpha^{\rm ref} \leqslant 427.12$~\cite{Gobre2016} and
$2.65 \leqslant R_{\rm vdW}^{\rm ref} \leqslant 6.24$. Both the mean of the relative error,
$\langle\text{R.E.}\rangle$, and its magnitude, $\langle|\text{R.E.}|\rangle$, are within a few percent.
Moreover, $\langle\text{R.E.}\rangle$ for the complete database of Batsanov is just 0.61\%,
which means that positive and negative deviations are almost equally distributed. Since the reference vdW
radii are determined with a statistical error of up to 10\%~\cite{Batsanov2001}, these results are already
enough to support the validity of Eq.~(\ref{eq.:Relation_2}).

The reliability of the obtained formula becomes even more evident from a further detailed analysis based
on our separate treatment of the nonmetals/metalloids and metals. The experimentally based determination
of $R_{\rm vdW}$ is known to be more difficult for atoms with metallic properties~\cite{Batsanov2001},
because of lack of structures where they undergo vdW-bonded contacts with other molecular moieties.
The transition elements are even more problematic since they exhibit a variety of possible electronic states.
Therefore, going from nonmetals via metalloids and simple metals to transition metals, the statistical
error increases. Figure 2 clearly demonstrates such a situation. On one hand, for the organic
elements (C, N, O) the agreement is better in comparison to the metalloids (As, Sb, Te). On the other hand,
the transition metals (V, Cr, Pd) show larger deviations in comparison to the simple metals (K, Rb, Sr).
It is also worth mentioning that, among all the elements from the used database~\cite{Comment_6},
$|\text{R.E.}|$ exceeds 10\% only for V, Cr, and Pd.

\begin{figure}[t!]
\includegraphics[width=0.7\LL]{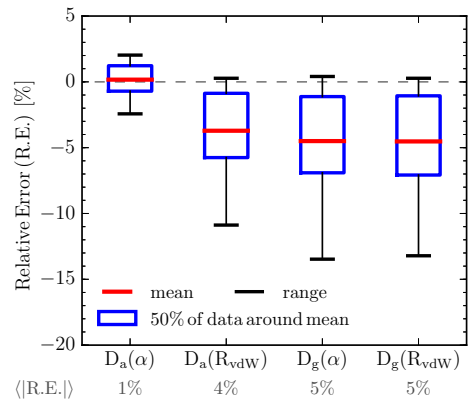}
\caption{(Color online)
Statistical analysis of the results obtained with Eqs.~(\ref{eq.:Re_alpha_A})--(\ref{eq.:Re_R_G}) for the equilibrium
distance of 15 vdW-bonded heteronuclear dimers of noble gases (all possible pairs among He, Ne, Ar, Kr, Xe, and Rn)
which is performed by comparison to the references values~\cite{Tang2003}.}
\label{img:Heteronuclear}
\end{figure}

An important feature of Eq.~(\ref{eq.:Relation_2}) is its transferability to vdW-bonded
heteronuclear dimers. The equilibrium distance between two different atoms $A$ and $B$ can be obtained by
the arithmetic mean
\begin{align}
\begin{array}{ll}
D_{\rm a} (\alpha) = 2 \times 2.54 \left[ (\alpha_A+\alpha_B)/2 \right]^{\nicefrac 17} \text{a.u.}
\end{array}
\label{eq.:Re_alpha_A}
\end{align}
as generalization of the equilibrium distance in homonuclear dimers,
$D (\alpha) \equiv 2 \times R_{\rm vdW} = 2 \times 2.54\,\alpha^{\nicefrac 17}$.
The box plot of Fig.~\ref{img:Heteronuclear} illustrates that the simple combination rule
of Eq.~(\ref{eq.:Re_alpha_A}) yields accurate equilibrium distances of 15 vdW-bonded heteronuclear
dimers of noble gases. The corresponding $|{\rm R.E.}|$ with respect to the reference data~\cite{Tang2003}
is within 2.5\%, whereas $\langle{\rm R.E.}\rangle$ and $\langle|{\rm R.E.}|\rangle$ are about 0.2\% and 1\%,
respectively~\cite{Supplementary}. In comparison, the other three possible combination rules based on simple means,
\begin{align}
\begin{array}{ll}
D_{\rm a} (R_{\rm vdW}) = 2 \times \left( R_{\rm vdW}^A + R_{\rm vdW}^B \right)/ 2\ ,
\end{array}
\label{eq.:Re_R_A}
\end{align}
\begin{align}
\begin{array}{ll}
D_{\rm g} (\alpha) = 2 \times 2.54 \left(\sqrt{\alpha_A\alpha_B} \right)^{\nicefrac 17}\ ,
\end{array}
\label{eq.:Re_alpha_G}
\end{align}
\begin{align}
\begin{array}{ll}
D_{\rm g} (R_{\rm vdW}) = 2 \times \left( R_{\rm vdW}^A R_{\rm vdW}^B \right)^{\nicefrac 12}\ ,
\end{array}
\label{eq.:Re_R_G}
\end{align}
underestimate the equilibrium distances with $|\text{R.E.}|$ exceeding 10\% and both $\langle{\rm R.E.}\rangle$
and $\langle|{\rm R.E.}|\rangle$ of about 4-5\%.

Among its various possible applications, the proposed determination of the atomic vdW radius and the equilibrium distance
for vdW bonds provides a powerful way to parametrize interatomic potentials. Many models, like the Lennard-Jones potential,
use a geometric and an energetic parameter. The former, related to the equilibrium distance,
can now be determined via the polarizability
according to Eqs.~(\ref{eq.:Relation_2}) and (\ref{eq.:Re_alpha_A}). Since the remaining parameter
corresponds to the dissociation energy, the full parametrization becomes now easily accessible by experiment. There are
also models, like the modified Tang-Toennies potential~\cite{Tang1995}, based just on one combined parameter, which can be
now directly evaluated from the extremum condition on the known equilibrium distance.

Based on Eq.~(\ref{eq.:Relation_2}), one can also significantly improve the efficiency of computational models
for intermolecular interactions by revising the determination of effective vdW radii of atoms in molecules.
According to the classical result, the vdW radius is conventionally calculated as
$R_{\rm vdW}^{\rm eff} = ({\alpha^{\rm eff}}/{\alpha^{\rm free}})^{\nicefrac 13} R_{\rm vdW}^{\rm free}$
with the effective atomic polarizability obtained from the corresponding electron density~\cite{Tkatchenko2009}.
To apply this procedure, it is necessary to tabulate empirical free-atom vdW radii.
With Eq.~(\ref{eq.:Relation_2}), this problem can now be overcome by a direct calculation
$R_{\rm vdW}^{\rm eff} = 2.54\,(\alpha^{\rm eff})^{\nicefrac 17}$. We test the effect of using
this alternative definition of vdW radii for atoms in molecules on the binding energies of molecular
dimers contained in the S66 database~\cite{Rezac2011_1} by means of the Tkatchenko-Scheffler model~\cite{Tkatchenko2009}
in conjunction with DFT-PBE calculations~\cite{Supplementary}. With the alternative determination
of $R_{\rm vdW}^{\rm eff}$ we obtain an accuracy increase of about 30\%, in comparison to the conventional and more empirical
computational scheme~\cite{Supplementary}. Hence, the use of Eq.~(\ref{eq.:Relation_2})
improves the accuracy of intermolecular interaction models as well as reduces their empiricism.

We have also found that Eq.~(\ref{eq.:Relation_2}) can be generalized to
\begin{align}
\begin{array}{ll}
R_{\rm vdW} (\alpha_n) = C_n~\alpha_n^{\nicefrac 2{7(n+1)}}\ \ ,\ \ \ n = 1, 2, ...
\end{array}
\label{eq.:Relation_general}
\end{align}
for the multipole polarizabilities~\cite{Comment_7}.
With the coefficients $C_2=2.45$ and $C_3=2.27$ as well as accurate values for $\alpha_2$ and $\alpha_3$ from
Ref.~\cite{Jones2013}, Eq.~(\ref{eq.:Relation_general}) provides $R_{\rm vdW}$ for He, Ne, Ar,
Kr, and Xe within 1\% and 1.4\%, respectively. This indicates that higher-order attractive and
repulsive forces related to each term in the Coulomb potential expansion are mutually balanced as well, which
justifies the model we used to derive the scaling law of Eq.~(\ref{eq.:Relation_2})~\cite{Comment_8}.

In summary, the present work provides a seamless and universal definition of the vdW radius for all chemical elements
solely in terms of their dipole polarizabilities, which is given by $R_{\rm vdW} (\alpha) = 2.54~\alpha^{\nicefrac 17}$.
Motivated by the definition of the vdW radius of Pauling~\cite{Pauling1960} and Bondi~\cite{Bondi1964},
this relation has been evaluated  by using the quantum Drude oscillator model for valence electronic response.
Notably, our finding implies a significant departure from the commonly employed classical scaling law,
$R_{\rm vdW} \propto \alpha^{\nicefrac 13}$. In-depth analysis of the most comprehensive empirical
reference radii~\cite{Batsanov2001} confirms the revealed quantum-mechanical relation. Our derivation of the vdW radius
dispenses with the need for its experimental determination. Moreover, the obtained relation is also successfully extended
to vdW-bonded heteronuclear dimers and higher-order atomic polarizabilities. The presented results
motivate future studies towards understanding the dependence of local geometric descriptors of an embedded atom on its
chemical environment as well as unveiling a non-trivial relationship between length and volume in quantum-mechanical
systems~\cite{Comment_9}.
\\ \\
We acknowledge financial support from the European Research Council (ERC Consolidator Grant \lq\lq BeStMo\rq\rq).
M.~St\"ohr acknowledges financial support from Fonds National de la Recherche, Luxembourg (AFR PhD Grant \lq\lq CNDTEC\rq\rq).
We are also thankful to Dr. Igor Poltavsky for valuable discussions.


\pagebreak


\onecolumngrid
\begin{center}
\phantom{}
\vskip 18cm
\textbf{\Large Quantum-Mechanical Relation between Atomic Dipole Polarizability \\ and the van der Waals Radius (Supplemental Material)}\\[.5cm]
{\large Dmitry~V.~Fedorov,$^{1,*}$ Mainak~Sadhukhan,$^{1}$ Martin~St\"ohr,$^{1}$ and Alexandre~Tkatchenko$^{1}$}\\[.5cm]
{\large \itshape ${}^1$Physics and Materials Science Research Unit, University of Luxembourg, L-1511 Luxembourg}\\[.2cm]
\end{center}

\setcounter{equation}{0}
\setcounter{figure}{0}
\setcounter{table}{0}
\setcounter{page}{1}
\renewcommand{\theequation}{S\arabic{equation}}
\renewcommand{\thefigure}{S\arabic{figure}}
\renewcommand{\thetable}{S\Roman{table}}
\renewcommand{\bibnumfmt}[1]{[S#1]}
\renewcommand{\citenumfont}[1]{S#1}

%
%
%
%
%

\Large

\newpage
\centerline{\Large \bf Derivation of the repulsive exchange energy within the QDO model}

\vskip 0.5cm
Here, the Heitler-London approach~[30] is applied to the quantum Drude oscillator model~[21]. For a homonuclear dimer
consisting of atoms $A$ and $B$ separated by $\mathbf{R}$, the corresponding atomic QDO wave functions are given by
$$
\Psi_A (\mathbf{r}) = \left(\frac{\mu\omega}{\pi\hbar}\right)^{\nicefrac 34} e^{-\frac{\mu\omega}{2\hbar}r^2}\ \ \text{and}
\ \ \Psi_B (\mathbf{r}) = \left(\frac{\mu\omega}{\pi\hbar}\right)^{\nicefrac 34} e^{-\frac{\mu\omega}{2\hbar}(\mathbf{r}-\mathbf{R})^2}\ ,
\eqno ({\rm S}1)
$$
respectively. The related overlap integral is
$$
S = \iint d \mathbf{r}_1 d \mathbf{r}_2~\Psi_A^* (\mathbf{r}_1) \Psi_B^* (\mathbf{r}_2)
\Psi_B (\mathbf{r}_1) \Psi_A (\mathbf{r}_2) = e^{-\frac{\mu\omega}{2\hbar}R^2}\ .
\eqno ({\rm S}2)
$$
We use the dipole approximation for the Coulomb interaction
$$
{\hat V}_{\rm dip} = q^2 \left\{ \frac{[\mathbf{r}_1 \cdot (\mathbf{r}_2 - \mathbf{R})]}{R^3} -
\frac{3 (\mathbf{r}_1 \cdot \mathbf{R}) [(\mathbf{r}_2 - \mathbf{R})\cdot \mathbf{R}]}{R^5} \right\}\ ,
\eqno ({\rm S}3)
$$
where the origin of the coordinates $\mathbf{r}_1$ and $\mathbf{r}_2$ of the two QDOs is located
on atom $A$.

Then, the corresponding Coulomb and exchange integrals are obtained as
$$
C = \iint d \mathbf{r}_1 d \mathbf{r}_2~\Psi_A^* (\mathbf{r}_1) \Psi_B^* (\mathbf{r}_2)
{\hat V}_{\rm dip} \Psi_A (\mathbf{r}_1) \Psi_B (\mathbf{r}_2) = 0
\eqno ({\rm S}4)
$$
and
$$
J_{\rm ex} = \iint d \mathbf{r}_1 d \mathbf{r}_2~\Psi_A^* (\mathbf{r}_1) \Psi_B^* (\mathbf{r}_2)
{\hat V}_{\rm dip} \Psi_B (\mathbf{r}_1) \Psi_A (\mathbf{r}_2) = \frac{q^2 S}{2R}\ ,
\eqno ({\rm S}5)
$$
respectively. We assume that the coarse-grained atomic QDO wave functions represent closed electronic shells
with total zero spin. According to their bosonic nature, the dimer wave function can only be symmetric (permanent)
$$
\Psi (\mathbf{r}_1, \mathbf{r}_2) = \frac 1{\sqrt{2}} \left[\Psi_A (\mathbf{r}_1)\Psi_B (\mathbf{r}_2) + \Psi_A (\mathbf{r}_2)\Psi_B (\mathbf{r}_1)\right]
\eqno ({\rm S}6)
$$
with the corresponding energy obtained as
$$
E = \frac{\iint d \mathbf{r}_1 d \mathbf{r}_2 \Psi^* (\mathbf{r}_1, \mathbf{r}_2) \hat{H} \Psi (\mathbf{r}_1, \mathbf{r}_2)}
{\iint d \mathbf{r}_1 d \mathbf{r}_2 \Psi^* (\mathbf{r}_1, \mathbf{r}_2) \Psi (\mathbf{r}_1, \mathbf{r}_2)}
= 2 E_0 + \frac{C + J_{\rm ex}}{1 + S} = 2 E_0 + \frac{J_{\rm ex}}{1 + S}\ ,
\eqno ({\rm S}7)
$$
where
$$
\hat{H}  = \hat{H}_0 (\mathbf{r}_1) + \hat{H}_0 (\mathbf{r}_2) + {\hat V}_{\rm dip} (\mathbf{r}_1, \mathbf{r}_2)
\eqno ({\rm S}8)
$$
with
$$
\hat{H}_0 (\mathbf{r}) \Psi_A (\mathbf{r}) = E_0 \Psi_A (\mathbf{r})\ \ \ \text{and}
\ \ \ \hat{H}_0 (\mathbf{r}) \Psi_B (\mathbf{r}) = E_0 \Psi_B (\mathbf{r})\ .
\eqno ({\rm S}9)
$$

At the equilibrium distance, $R = 2 R_{\rm vdW}$, of homonuclear dimers consisting of the species of Table~I,
the condition $S \ll 1$ is fulfilled:

\hskip 7cm $S_{\text{\tiny He--He}} = 6.94 \times 10^{-4}$\ ,

\hskip 7cm $S_{\text{\tiny Ne--Ne}} = 4.69 \times 10^{-4}$\ ,

\hskip 7cm $S_{\text{\tiny Ar--Ar}} = 3.95 \times 10^{-3}$\ ,

\hskip 7cm $S_{\text{\tiny Kr--Kr}} = 5.58 \times 10^{-3}$\ ,

\hskip 7cm $S_{\text{\tiny Xe--Xe}} = 1.28 \times 10^{-2}$\ ,

\hskip 7cm $S_{\text{\tiny Rn--Rn}} = 2.01 \times 10^{-2}$\ .

\vskip 0.3cm
\noindent Then, neglecting the second and higher order terms
with respect to the overlap integral, the repulsive exchange energy can be well approximated by

$$
E_{\rm ex} \approx J_{\rm ex} = \frac{q^2 S}{2R} = \frac{q^2}{2R}~e^{-\frac{\mu\omega}{2\hbar} R^2}\ .
\eqno ({\rm S}10)
$$

\noindent The corresponding force is obtained as

$$
F_{\rm ex} = - \nabla_R J_{\rm ex} = \frac{q^2}{2} \left[ \frac{\mu\omega}{\hbar} +
\frac{1}{R^2} \right]~e^{-\frac{\mu\omega}{2\hbar} R^2}\ .
\eqno ({\rm S}11)
$$

\vskip 0.2cm
\noindent As follows from Table~I, for $R = 2 R_{\rm vdW}$, the condition $\frac{\mu\omega}{\hbar} \gg \frac{1}{R^2}$
is fulfilled. Then, one can use the following approximation

$$
F_{\rm ex} \approx \frac{q^2}{2} \frac{\mu\omega}{\hbar}~e^{-\frac{\mu\omega}{2\hbar} R^2}
= \frac{\alpha \hbar\omega}{2} \left(\frac{\mu\omega}{\hbar}\right)^2~e^{-\frac{\mu\omega}{2\hbar} R^2}\ ,
\eqno ({\rm S}12)
$$

\vskip 0.2cm
\noindent where it is taken into account that the dipole polarizability is given by $\alpha = q^2 / \mu \omega^2$
within the QDO model~[21].

\vskip 0.4cm
The performed derivation is not as obvious as the more conventional approach~[28] efficiently used for noble gases, which
is based on the consideration of each single pair of interacting electrons. Such a detailed treatment is impossible within
the coarse-grained QDO model~[21], where the wave function of a single oscillator (a Drude particle) represents all valence
electrons together. However, taking into account the bosonic nature of closed valence electron shells, our approach is straightforward.
The validity of Eqs.~(S10) and (S12) is confirmed by the reasonable agreement between the ratio $R_{\rm vdW}/\alpha^{\nicefrac 17}$
obtained either within the QDO model or with the reference data for real atoms, as demonstrated by Table~I.

\vskip 4cm
\centerline{\Large \bf Extended statistical analysis for the noble gases}

\vskip 0.5cm
Here, we perform an extended statistical analysis for the six noble gases of Table~I, considering the function
$$
C (p, \alpha, R_{\rm vdW}) = R_{\rm vdW}^{\rm ref} / (\alpha^{\rm ref})^p\ .
\eqno ({\rm S}13)
$$
The reference values for the atomic dipole polarizability and the vdW radius are taken from Refs.~[12] and [8,33] of the main
manuscript, respectively.
The following different possible power laws are assumed: $p \in \{ {\nicefrac 13}; {\nicefrac 14}; {\nicefrac 15}; {\nicefrac 16};
{\nicefrac 17}; {\nicefrac 18}; {\nicefrac 19}; {\nicefrac 1{10}}; {\nicefrac 1{100}} \}$.
We calculate the arithmetic mean
$$
\langle C \rangle = \frac 16 \sum\limits_{i=1}^{6} C_i\ \ ,
\eqno ({\rm S}14)
$$
the standard deviation

$$
\sigma = \left[ \frac 1{(6-1)} \sum\limits_{i=1}^{6} \left(C_i - \langle C \rangle\right)^2 \right]^{\nicefrac 12}\ ,
\eqno ({\rm S}15)
$$

\vskip 0.3cm
\noindent and the coefficient of variation

$$
c_v = \frac{\sigma}{\langle C \rangle} \times 100\%\ \ .
\eqno ({\rm S}16)
$$

\vskip 0.4cm
\noindent The corresponding results are shown in Table~S\,I. Obviously, the relation between the vdW radius and the polarizability
given by Eq.~(7) is most reasonable among all considered power laws. The related standard deviation of 0.02 is the minimal one
and the coefficient of variation is less than 1\%. The other assumed relations do not provide such a good statistical picture.
These results serve as an additional confirmation of the obtained scaling law
solely from the statistical analysis of the reference data.

\vskip 0.6cm
\begin{table}[h!]
\footnotesize
\caption{\footnotesize  Results of the extended statistical analysis for the noble gases (He, Ne, Ar, Kr, Xe, and Rn).}
\begin{center}
\begin{tabular}{|c||c|c|c|c|c|c|c|c|c|}
\hline
$p$ & $\nicefrac 13$ & $\nicefrac 14$ & $\nicefrac 15$ & $\nicefrac 16$ & $\mathbf{\nicefrac 17}$ & $\nicefrac 18$ & $\nicefrac 19$ & $\nicefrac 1{10}$ & $\nicefrac 1{100}$ \\
\hline
~~$\langle C \rangle$~~ & 1.71 & 2.02 & 2.24 & 2.41 & {\bf 2.54} & 2.64 & 2.73 & 2.80 & 3.46 \\
\hline
$\sigma$ & 0.44 & 0.28 & 0.16 & 0.07 & {\bf 0.02} & 0.07 & 0.12 & 0.17 & 0.58 \\
\hline
$c_v$ & ~~25.51\%~~ & ~~13.92\%~~ & ~~~7.15\%~~ & ~~~2.75\%~~ & ~~{\bf ~0.65\%}~~ & ~~~2.75\%~~ & ~~~4.50\%~~ & ~~~5.90\%~~ & ~~16.85\%~~ \\
\hline
\end{tabular}
\end{center}
\end{table}

\vskip 4cm
\centerline{\Large \bf Dependence of the obtained results on the reference polarizability}

\vskip 0.5cm
In the main manuscript, we used the atomic dipole polarizability from Table A.1 of Ref.~[12], as the reference dataset.
Here, Fig.~S\,1 shows the results obtained with $\alpha$ taken from the benchmark \lq\lq Dataset for All Neutral Atoms\rq\rq\ 
of Ref.~[14]. Comparing it to Fig.~2, the mean of the relative error, $\langle {\rm R.E.}\rangle$, and its magnitude,
$\langle |{\rm R.E.}|\rangle$, are pratically the same. This is caused by the fact that $\alpha$ is a well-determined
quantity [9-14]. A remarkable difference is present only for Pd where R.E. changes from -10.11\% to 3.07\%. However, as discussed
in the main manuscript, the values of $R_{\rm vdW}^{\rm ref}$ for transition metals are not well reliable, to judge which from
the two polarizabilities for Pd is more accurate. Moreover, for organic elements corresponding to the most robust values of
$R_{\rm vdW}^{\rm ref}$, the agreement with the reference vdW radius becomes slightly worse in comparison to Fig.~2. In addition
to Fig.~S\,1, Table~S\,II provides a more detailed information about the used reference data as well as the results obtained for
$R_{\rm vdW} (\alpha)$.
\begin{figure}[h!]
\includegraphics[width=0.62\LL]{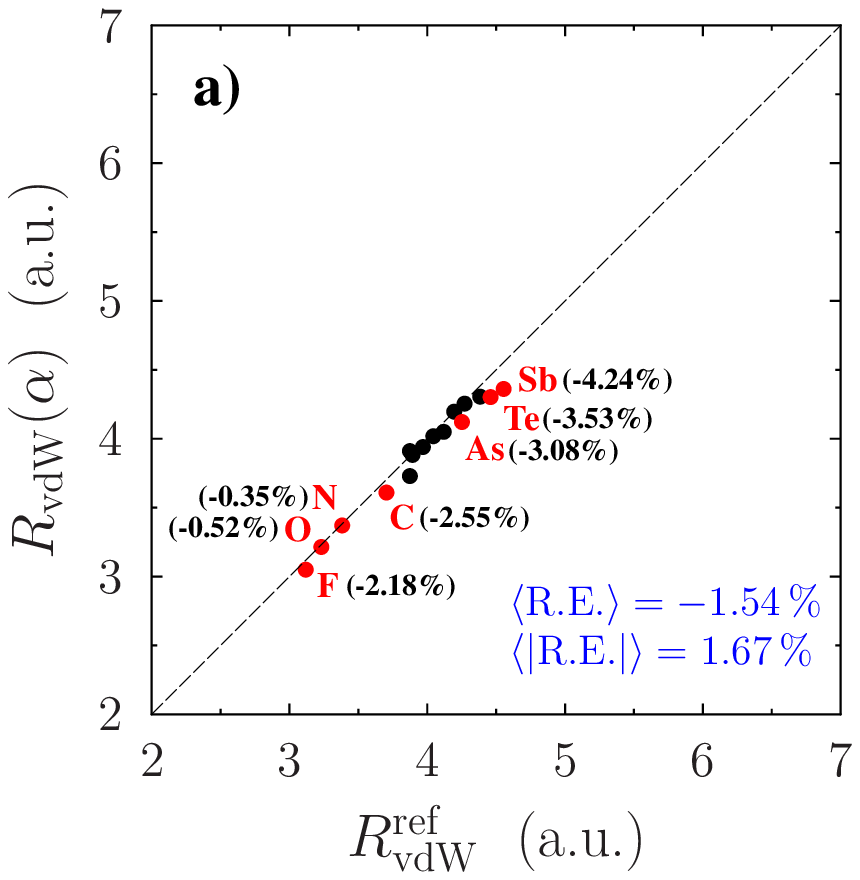}
\hskip 1.2cm
\includegraphics[width=0.62\LL]{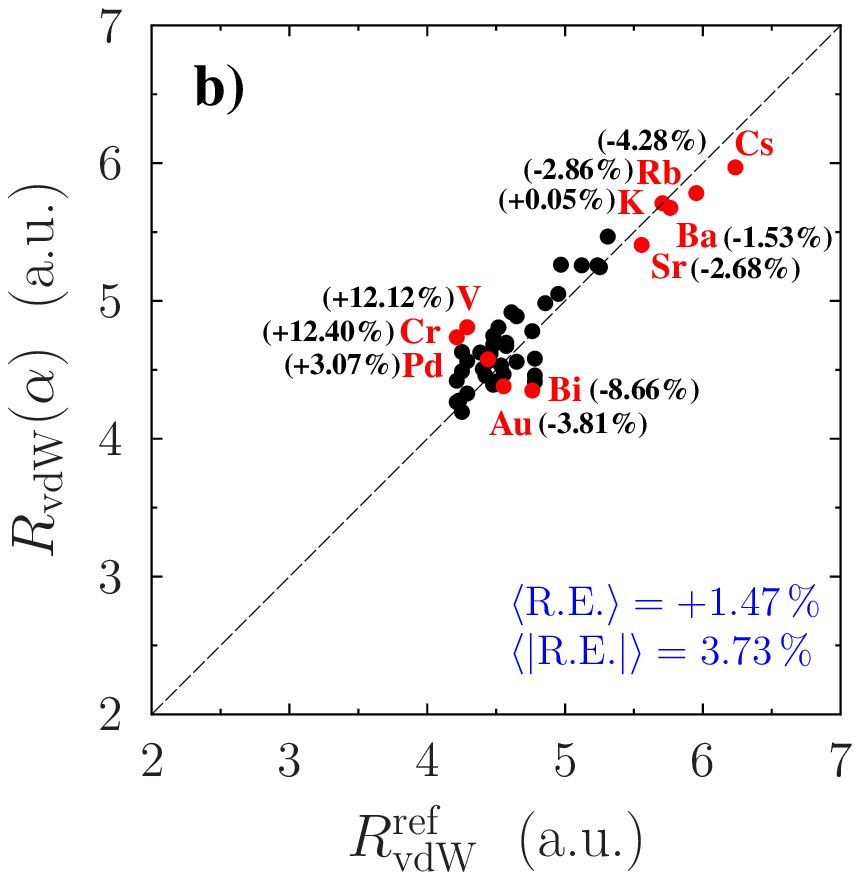}
\caption{(Color online) The vdW radius obtained with Eq.~(7) by using the reference data for
the polarizability from Ref.~[14] is shown separately for a) nonmetals/metalloids and b) metals in comparison
to its reference counterpart~[15]. Here, $\langle\text{R.E.}\rangle$ and $\langle|\text{R.E.}|\rangle$
represent the mean of the relative error and its magnitude, respectively, calculated with Eq.~(8) taking
$R_{\rm vdW}^{\rm ref}$ from the database of Batsanov [16].}
\label{fig.:S1}
\end{figure}

Based on the analysis of the obtained results, we conclude that the choice of the reference dipole polarizability
between two available databases plays no role for our conclusions made in the main manuscript.

\vskip 5cm
\centerline{\Large \bf Equilibrium distance in heteronuclear dimers of noble gases}

\vskip 0.5cm
As complementary to Fig.~3 of the main manuscript, Fig.~S\,2 and Table~S\,III present more detailed results for the equilibrium
distance in vdW-bonded heteronuclear dimers of noble gases obtained with Eqs.~(9)--(12). In principle, Eqs.~(11) and (12) are
equivalent, since
$$
2.54\, (\sqrt{\alpha_A \alpha_B})^{\nicefrac 17} = \sqrt{R_{\rm vdW}^A R_{\rm vdW}^B}\ ,
\eqno ({\rm S}17)
$$

\vskip 0.15cm
\noindent according to Eq.~(7). The present tiny differences between the related results are caused by errors in the reference
data for the vdW radii and the polarizabilities. In comparison to Eqs.~(11) and (12), the results obtained with Eq.~(10)
are slightly more accurate. Taking into account that Eqs.~(10) and (12) are the two approximations often used in literature,
we can judge that the approach based on the arithmetic mean for the vdW radii is preferable. The related formula can be used
for reasonable estimations of the equilibrium distance providing the relative error within 10\%. However, Eq.~(9), as
the generalization of Eq.~(7), provides much more accurate results with ${\rm R.E.}$ within 2.5\%,
$\langle {\rm R.E.} \rangle$ = 0.2\% and $\langle |{\rm R.E.}| \rangle = 1\%$.
\begin{figure}[h!]
\includegraphics[width=0.78\LL]{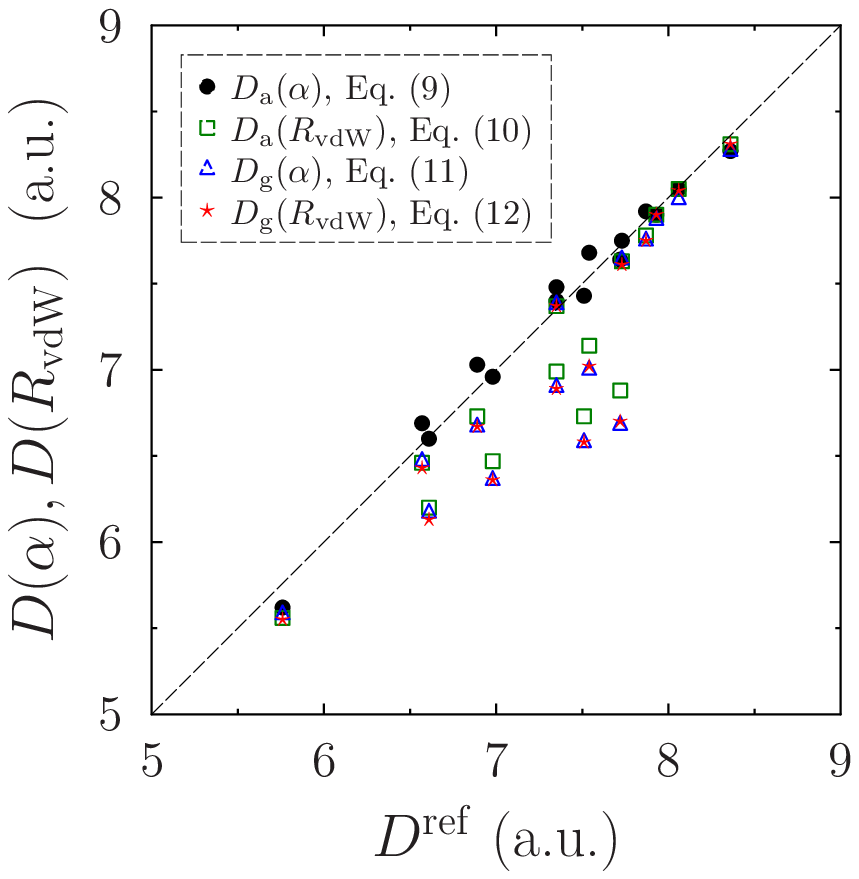}
\caption{(Color online) The equilibrium distance of 15 vdW-bonded heteronuclear dimers of the noble gases
(all possible pairs between He, Ne, Ar, Kr, Xe, and Rn) calculated with Eqs.~(9)--(12) versus its reference counterpart~[29].}
\label{fig.:S2}
\end{figure}

\vskip 5cm
\centerline{\Large \bf Calculation of the effective atomic vdW radius in molecules}

\vskip 0.5cm
Here, we present detailed numerical results of our test calculations performed for molecular dimers from the S66 database~[37]
by means of the Tkatchenko-Scheffler (TS) method~[10] with the Perdew-Burke-Ernzerhof (PBE) exchange-correlation functional~[43].
The absolute and relative error of the interaction energy with respect to the coupled-cluster reference data~[38] are obtained
either with the old

$$
R_{\rm vdW}^{\rm eff} = \left( \alpha^{\rm eff} / \alpha^{\rm free} \right)^{\nicefrac 13} R_{\rm vdW}^{\rm free}
= \left( V^{\rm eff} / V^{\rm free} \right)^{\nicefrac 13} R_{\rm vdW}^{\rm free}
\eqno ({\rm S}18)
$$

\noindent or new

$$
R_{\rm vdW}^{\rm eff} = 2.54 \left( \alpha^{\rm eff} \right)^{\nicefrac 17}
\eqno ({\rm S}19)
$$

\vskip 0.6cm
\noindent way to calculate the effective atomic vdW radius in molecules. Here, the polarizability/volume ratio is obtained by means of
the Hirshfeld partitioning of the electron density~[10].

Similar to the approach of Ref.~[10], we refitted the TS damping parameter $s_R$ employing Eq.~(S19) on the S22 benchmark set
of molecular dimers~[44]. With its new value of $0.91$, in comparison to $0.94$~[10] obtained with the old scheme based on
Eq.~(S18), we perform our test calculations on the S66 dataset~[37,38]. The corresponding density-functional calculations have
been carried out using the all-electron code FHI-aims~[45] with \emph{tight} defaults.

As demonstrated by Table~S\,IV, the new approach provides better accuracy for 57 from 66 molecular dimers. For nine other systems,
the difference between the absolute errors related to the two approaches is still much less than the chemical accuracy (1 kcal/mol).
Therefore, we may conclude that Eq.~(S19) provides a general improvement of the used procedure. The mean relative error and its
magnitude change from $-$11.5\% to $-$7.4\% and from 12.0\% to 8.8\%, respectively, by using the new approach instead of the old one.
This corresponds to the accuracy improvement by about 30\%.

This finding provides an additional confirmation of the revealed scaling law pointing out to a quite non-trivial
quantum-mechanical relation between the atomic volume and the vdW radius.

\vskip 5cm
\centerline{\Large \bf Symmetry-adapted perturbation theory analysis}

\vskip 0.5cm
With the symmetry-adapted perturbation theory (SAPT) decomposition, one obtains six contributions: electrostatic
($E_{\rm pol}^{(1)}$), exchange ($E_{\rm exch}^{(1)}$), induction ($E_{\rm ind}^{(2)}$), exchange-induction
($E_{\rm exch-ind}^{(2)}$), dispersion ($E_{\rm disp}^{(2)}$), and exchange-dispersion ($E_{\rm exch-disp}^{(2)}$).
In the case of neutral systems, the net induction interaction
($E_{\rm ind}^{(2)} + E_{\rm exch-ind}^{(2)}$) is almost zero due to the balance between its constituents~[46]
and the decomposition can be restricted to the four other contributions.
For noble gas dimers, numerical results of the SAPT based on coupled-cluster
approach with single and double excitations (CCSD) were provided recently in Ref.~[47]. The authors have shown that
their calculations are in good agreement with corresponding density-functional theory (DFT) based SAPT approaches.
By using their data, we evaluate the SAPT contributions to attractive and repulsive forces for He-He, Ne-Ne, Ar-Ar,
and Kr-Kr dimers considered in Ref.~[47]. The magnitude of corresponding contributions is obtained within the following
spans: $0.15 < F_{\rm pol}^{(1)}/F_{\rm exch}^{(1)} < 0.35$, $0.65 < F_{\rm disp}^{(2)}/F_{\rm exch}^{(1)} < 0.82$,
and  $0.05 < F_{\rm exch-disp}^{(2)}/F_{\rm exch}^{(1)} < 0.13$, where the maximal contribution $F_{\rm exch}^{(1)}$ is
chosen as a reference. The net induction force ($F_{\rm ind}^{(2)} + F_{\rm exch-ind}^{(2)}$) is one order of magnitude
less than $F_{\rm exch-disp}^{(2)}$ and therefore can be disregarded. This analysis shows that the force stemming from
the electrostatic interaction has a relevant contribution. However, by its definition (for instance, Eq.~(1) in
Ref.~[48]), $E_{\rm pol}^{(1)}$ is equal to the Coulomb integral which, like our Eq.~(S4), vanishes in the dipole
approximation for spherically symmetric atomic electron densities.
Therefore, the corresponding force can contribute only for higher-order terms in multipole expansion of the Coulomb
potential. Its influence needs to be considered for derivation of the general relationship given by Eq.~(13), but it
is irrelevant for our derivation of the scaling law expressed by Eq.~(7).

\begin{table}[h!]
\scriptsize
\caption{\footnotesize The van der Waals radius calculated according to Eq.~(7) with the atomic dipole polarizability taken
either from Ref.~[12] ($\alpha^{(1)}$) or from Ref.~[14] ($\alpha^{(2)}$) is presented in comparison to its reference [15] counterpart.
The relative error (${\rm R.E.}$) is calculated by Eq. (8). A comment: for Th and U, there are no data in Ref.~[14].}
\begin{center}
\begin{tabular}{|c||c|c|c|c|c|c|c|}
\hline
Species & $R_{\rm vdW}^{\rm ref}$ &  $\alpha^{(1)}$ & $R_{\rm vdW} (\alpha^{(1)})$ & ${\rm R.E.}$ & $\alpha^{(2)}$ & $R_{\rm vdW} (\alpha^{(2)})$ & ${\rm R.E.}$ \\
\hline
Li   &   ~~4.9700~~  &  ~~164.2000~~ &   5.2640   &  ~~$+$5.92\%~~    &  ~~164.00~~  &   5.2631   &  ~~$+$5.90\%~~    \\
Be   &     4.2141    &    ~38.0000   &   4.2709   &  ~~$+$1.35\%~~    &    ~37.70    &   4.2660   &  ~~$+$1.23\%~~    \\
 B   &     3.8739    &    ~21.0000   &   3.9239   &  ~~$+$1.29\%~~    &    ~20.50    &   3.9105   &  ~~$+$0.94\%~~    \\
 C   &     3.7039    &    ~12.0000   &   3.6225   &  ~~$-$2.20\%~~    &    ~11.70    &   3.6094   &  ~~$-$2.55\%~~    \\
 N   &     3.3826    &    ~~7.4000   &   3.3807   &  ~~$-$0.06\%~~    &    ~~7.25    &   3.3709   &  ~~$-$0.35\%~~    \\
 O   &     3.2314    &    ~~5.4000   &   3.2319   &  ~~$+$0.02\%~~    &    ~~5.20    &   3.2146   &  ~~$-$0.52\%~~    \\
 F   &     3.1180    &    ~~3.8000   &   3.0737   &  ~~$-$1.42\%~~    &    ~~3.60    &   3.0500   &  ~~$-$2.18\%~~    \\
Na   &     5.2345    &    162.7000   &   5.2571   &  ~~$+$0.43\%~~    &    163.00    &   5.2585   &  ~~$+$0.46\%~~    \\
Mg   &     4.5731    &    ~71.0000   &   4.6698   &  ~~$+$2.11\%~~    &    ~71.40    &   4.6736   &  ~~$+$2.20\%~~    \\
Al   &     4.5353    &    ~60.0000   &   4.5589   &  ~~$+$0.52\%~~    &    ~57.50    &   4.5312   &  ~~$-$0.09\%~~    \\
Si   &     4.2708    &    ~37.0000   &   4.2546   &  ~~$-$0.38\%~~    &    ~37.00    &   4.2546   &  ~~$-$0.38\%~~    \\
 P   &     4.0440    &    ~25.0000   &   4.0229   &  ~~$-$0.52\%~~    &    ~24.80    &   4.0183   &  ~~$-$0.64\%~~    \\
 S   &     3.8928    &    ~19.6000   &   3.8855   &  ~~$-$0.19\%~~    &    ~19.50    &   3.8826   &  ~~$-$0.26\%~~    \\
Cl   &     3.8739    &    ~15.0000   &   3.7398   &  ~~$-$3.46\%~~    &    ~14.70    &   3.7290   &  ~~$-$3.74\%~~    \\
 K   &     5.7070    &    292.9000   &   5.7177   &  ~~$+$0.19\%~~    &    290.00    &   5.7096   &  ~~$+$0.05\%~~    \\
Ca   &     5.2534    &    160.0000   &   5.2446   &  ~~$-$0.17\%~~    &    160.00    &   5.2446   &  ~~$-$0.17\%~~    \\
Sc   &     4.9511    &    120.0000   &   5.0334   &  ~~$+$1.66\%~~    &    123.00    &   5.0512   &  ~~$+$2.02\%~~    \\
Ti   &     4.6109    &    ~98.0000   &   4.8898   &  ~~$+$6.05\%~~    &    102.00    &   4.9179   &  ~~$+$6.66\%~~    \\
 V   &     4.2897    &    ~84.0000   &   4.7833   &  ~$+$11.51\%~~    &    ~87.30    &   4.8098   &  ~$+$12.12\%~~    \\
Cr   &     4.2141    &    ~78.0000   &   4.7330   &  ~$+$12.31\%~~    &    ~78.40    &   4.7364   &  ~$+$12.40\%~~    \\
Mn   &     4.2519    &    ~63.0000   &   4.5907   &  ~~$+$7.97\%~~    &    ~66.80    &   4.6293   &  ~~$+$8.88\%~~    \\
Fe   &     4.2897    &    ~56.0000   &   4.5141   &  ~~$+$5.23\%~~    &    ~60.40    &   4.5632   &  ~~$+$6.38\%~~    \\
Co   &     4.2519    &    ~50.0000   &   4.4416   &  ~~$+$4.46\%~~    &    ~53.90    &   4.4896   &  ~~$+$5.59\%~~    \\
Ni   &     4.2141    &    ~48.0000   &   4.4158   &  ~~$+$4.79\%~~    &    ~48.40    &   4.4211   &  ~~$+$4.91\%~~    \\
Cu   &     4.2897    &    ~42.0000   &   4.3324   &  ~~$+$1.00\%~~    &    ~41.70    &   4.3280   &  ~~$+$0.89\%~~    \\
Zn   &     4.2330    &    ~40.0000   &   4.3023   &  ~~$+$1.64\%~~    &    ~38.40    &   4.2773   &  ~~$+$1.05\%~~    \\
Ga   &     4.5542    &    ~60.0000   &   4.5589   &  ~~$+$0.10\%~~    &    ~52.10    &   4.4678   &  ~~$-$1.90\%~~    \\
Ge   &     4.3842    &    ~41.0000   &   4.3175   &  ~~$-$1.52\%~~    &    ~40.20    &   4.3054   &  ~~$-$1.80\%~~    \\
As   &     4.2519    &    ~29.0000   &   4.1091   &  ~~$-$3.36\%~~    &    ~29.60    &   4.1212   &  ~~$-$3.08\%~~    \\
Se   &     4.1196    &    ~25.0000   &   4.0229   &  ~~$-$2.35\%~~    &    ~26.20    &   4.0499   &  ~~$-$1.69\%~~    \\
Br   &     3.9684    &    ~20.0000   &   3.8967   &  ~~$-$1.81\%~~    &    ~21.60    &   3.9398   &  ~~$-$0.72\%~~    \\
Rb   &     5.9526    &    319.2000   &   5.7884   &  ~~$-$2.76\%~~    &    317.00    &   5.7827   &  ~~$-$2.86\%~~    \\
Sr   &     5.5558    &    199.0000   &   5.4106   &  ~~$-$2.61\%~~    &    198.00    &   5.4067   &  ~~$-$2.68\%~~    \\
 Y   &     5.1212    &    126.7370   &   5.0728   &  ~~$-$0.94\%~~    &    163.00    &   5.2585   &  ~~$+$2.68\%~~    \\
Zr   &     4.8566    &    119.9700   &   5.0332   &  ~~$+$3.64\%~~    &    112.00    &   4.9840   &  ~~$+$2.62\%~~    \\
Nb   &     4.6487    &    101.6030   &   4.9151   &  ~~$+$5.73\%~~    &    ~97.90    &   4.8891   &  ~~$+$5.17\%~~    \\
Mo   &     4.5164    &    ~88.4225   &   4.8185   &  ~~$+$6.69\%~~    &    ~87.10    &   4.8082   &  ~~$+$6.46\%~~    \\
Tc   &     4.4787    &    ~80.0830   &   4.7508   &  ~~$+$6.08\%~~    &    ~79.60    &   4.7467   &  ~~$+$5.99\%~~    \\
Ru   &     4.4787    &    ~65.8950   &   4.6203   &  ~~$+$3.16\%~~    &    ~72.30    &   4.6819   &  ~~$+$4.54\%~~    \\
Rh   &     4.3842    &    ~56.1000   &   4.5153   &  ~~$+$2.99\%~~    &    ~66.40    &   4.6253   &  ~~$+$5.50\%~~    \\
Pd   &     4.4409    &    ~23.6800   &   3.9919   &  ~$-$10.11\%~~    &    ~61.70    &   4.5771   &  ~~$+$3.07\%~~    \\
Ag   &   ~~4.4787~~  &  ~~~50.6000~~ &   4.4492   &  ~~$-$0.66\%~~    &    ~46.20    &   4.3918   &  ~~$-$1.94\%~~    \\
Cd   &     4.4787    &    ~39.7000   &   4.2977   &  ~~$-$4.04\%~~    &    ~46.70    &   4.3985   &  ~~$-$1.79\%~~    \\
In   &     4.7810    &    ~70.2200   &   4.6625   &  ~~$-$2.48\%~~    &    ~62.10    &   4.5813   &  ~~$-$4.18\%~~    \\
Sn   &     4.6487    &    ~55.9500   &   4.5136   &  ~~$-$2.91\%~~    &    ~60.00    &   4.5589   &  ~~$-$1.93\%~~    \\
Sb   &     4.5542    &    ~43.6719   &   4.3566   &  ~~$-$4.34\%~~    &    ~44.00    &   4.3613   &  ~~$-$4.24\%~~    \\
Te   &     4.4598    &    ~37.6500   &   4.2652   &  ~~$-$4.36\%~~    &    ~40.00    &   4.3023   &  ~~$-$3.53\%~~    \\
 I   &     4.1952    &    ~35.0000   &   4.2210   &  ~~$+$0.62\%~~    &    ~33.60    &   4.1965   &  ~~$+$0.03\%~~    \\
Cs   &     6.2361    &    427.1200   &   6.0343   &  ~~$-$3.24\%~~    &  ~~396.00~~  &   5.9694   &  ~~$-$4.28\%~~    \\
Ba   &     5.7637    &    275.0000   &   5.6664   &  ~~$-$1.69\%~~    &    278.00    &   5.6752   &  ~~$-$1.53\%~~    \\
La   &     5.3101    &    213.7000   &   5.4659   &  ~~$+$2.93\%~~    &    214.00    &   5.4670   &  ~~$+$2.95\%~~    \\
Hf   &     4.7621    &    ~99.5200   &   4.9006   &  ~~$+$2.91\%~~    &    ~83.70    &   4.7809   &  ~~$+$0.39\%~~    \\
Ta   &     4.5731    &    ~82.5300   &   4.7713   &  ~~$+$4.33\%~~    &    ~73.90    &   4.6966   &  ~~$+$2.70\%~~    \\
 W   &     4.4598    &    ~71.0410   &   4.6702   &  ~~$+$4.72\%~~    &    ~65.80    &   4.6193   &  ~~$+$3.58\%~~    \\
Re   &     4.4409    &    ~63.0400   &   4.5912   &  ~~$+$3.38\%~~    &    ~60.20    &   4.5610   &  ~~$+$2.71\%~~    \\
Os   &     4.4031    &    ~55.0550   &   4.5032   &  ~~$+$2.27\%~~    &    ~55.30    &   4.5060   &  ~~$+$2.34\%~~    \\
Ir   &     4.4220    &    ~42.5100   &   4.3399   &  ~~$-$1.86\%~~    &    ~51.30    &   4.4580   &  ~~$+$0.81\%~~    \\
Pt   &     4.4787    &    ~39.6800   &   4.2974   &  ~~$-$4.05\%~~    &    ~48.00    &   4.4158   &  ~~$-$1.40\%~~    \\
Au   &     4.5542    &    ~36.5000   &   4.2464   &  ~~$-$6.76\%~~    &    ~45.40    &   4.3808   &  ~~$-$3.81\%~~    \\
Hg   &     4.2519    &    ~33.9000   &   4.2018   &  ~~$-$1.18\%~~    &    ~33.50    &   4.1947   &  ~~$-$1.35\%~~    \\
Tl   &     4.7810    &    ~69.9200   &   4.6596   &  ~~$-$2.54\%~~    &    ~51.40    &   4.4592   &  ~~$-$6.73\%~~    \\
Pb   &     4.7810    &    ~61.8000   &   4.5781   &  ~~$-$4.24\%~~    &    ~47.90    &   4.4145   &  ~~$-$7.67\%~~    \\
Bi   &     4.7621    &    ~49.0200   &   4.4291   &  ~~$-$6.99\%~~    &    ~43.20    &   4.3499   &  ~~$-$8.66\%~~    \\
Th   &     5.1967    &    217.0000   &   5.4779   &  ~~$+$5.41\%~~    &      ---     &    ---     &      ---       \\
 U   &     5.0078    &    127.8000   &   5.0789   &  ~~$+$1.42\%~~    &      ---     &    ---     &      ---       \\
\hline
\end{tabular}
\end{center}
\end{table}

\begin{table}[t!]
\scriptsize
  \caption{\footnotesize The equilibrium distance of 15 vdW-bonded heteronuclear dimers of the noble gases
   (all possible pairs between He, Ne, Ar, Kr, Xe, and Rn) calculated with Eqs.~(9)--(12) in comparison to the reference values~[29].}
\begin{tabular}{|c||c|c|c|c|c|c|c|c|c|}
\hline
Dimer & $D^{\rm ref}$ & $D_{\rm a} (\alpha)$ & R.E. & $D_{\rm a} (R_{\rm vdW})$ & R.E. &
$D_{\rm g} (\alpha)$ & R.E. & $D_{\rm g} (R_{\rm vdW})$ & R.E. \\
\hline
~~He-Ne~~  &  ~~5.76~~  &  ~~5.62~~  &  ~~$-$2.43\%~~~  &  5.56  &   ~~$-$3.47\%~~~  &   5.58  &   ~~$-$3.13\%~~~  &  5.55  &   ~~$-$3.58\%~~  \\
~~He-Ar~~  &  ~~6.61~~  &  ~~6.60~~  &  ~~$-$0.15\%~~~  &  6.20  &   ~~$-$6.20\%~~~  &   6.17  &   ~~$-$6.66\%~~~  &  6.13  &   ~~$-$7.20\%~~  \\
~~He-Kr~~  &  ~~6.98~~  &  ~~6.96~~  &  ~~$-$0.29\%~~~  &  6.47  &   ~~$-$7.31\%~~~  &   6.36  &   ~~$-$8.88\%~~~  &  6.36  &   ~~$-$8.83\%~~  \\
~~He-Xe~~  &  ~~7.51~~  &  ~~7.43~~  &  ~~$-$1.07\%~~~  &  6.73  &   ~$-$10.39\%~~~  &   6.58  &   ~$-$12.38\%~~~  &  6.58  &   ~$-$12.43\%~~  \\
~~He-Rn~~  &  ~~7.72~~  &  ~~7.64~~  &  ~~$-$1.04\%~~~  &  6.88  &   ~$-$10.88\%~~~  &   6.68  &   ~$-$13.47\%~~~  &  6.70  &   ~$-$13.26\%~~  \\
~~Ne-Ar~~  &  ~~6.57~~  &  ~~6.69~~  &  ~~$+$1.83\%~~~  &  6.46  &   ~~$-$1.67\%~~~  &   6.47  &   ~~$-$1.52\%~~~  &  6.43  &   ~~$-$2.16\%~~  \\
~~Ne-Kr~~  &  ~~6.89~~  &  ~~7.03~~  &  ~~$+$2.03\%~~~  &  6.73  &   ~~$-$2.32\%~~~  &   6.67  &   ~~$-$3.19\%~~~  &  6.67  &   ~~$-$3.22\%~~  \\
~~Ne-Xe~~  &  ~~7.35~~  &  ~~7.48~~  &  ~~$+$1.77\%~~~  &  6.99  &   ~~$-$4.90\%~~~  &   6.90  &   ~~$-$6.12\%~~~  &  6.89  &   ~~$-$6.24\%~~  \\
~~Ne-Rn~~  &  ~~7.54~~  &  ~~7.68~~  &  ~~$+$1.86\%~~~  &  7.14  &   ~~$-$5.31\%~~~  &   7.00  &   ~~$-$7.16\%~~~  &  7.02  &   ~~$-$6.94\%~~  \\
~~Ar-Kr~~  &  ~~7.35~~  &  ~~7.40~~  &  ~~$+$0.68\%~~~  &  7.37  &   ~~$+$0.27\%~~~  &   7.38  &   ~~$+$0.41\%~~~  &  7.37  &   ~~$+$0.20\%~~  \\
~~Ar-Xe~~  &  ~~7.73~~  &  ~~7.75~~  &  ~~$+$0.26\%~~~  &  7.63  &   ~~$-$1.29\%~~~  &   7.64  &   ~~$-$1.16\%~~~  &  7.61  &   ~~$-$1.53\%~~  \\
~~Ar-Rn~~  &  ~~7.87~~  &  ~~7.92~~  &  ~~$+$0.64\%~~~  &  7.78  &   ~~$-$1.14\%~~~  &   7.75  &   ~~$-$1.52\%~~~  &  7.75  &   ~~$-$1.52\%~~  \\
~~Kr-Xe~~  &  ~~7.93~~  &  ~~7.90~~  &  ~~$-$0.38\%~~~  &  7.90  &   ~~$-$0.38\%~~~  &   7.87  &   ~~$-$0.76\%~~~  &  7.90  &   ~~$-$0.43\%~~  \\
~~Kr-Rn~~  &  ~~8.06~~  &  ~~8.05~~  &  ~~$-$0.12\%~~~  &  8.05  &   ~~$-$0.12\%~~~  &   7.99  &   ~~$-$0.87\%~~~  &  8.04  &   ~~$-$0.25\%~~  \\
~~Xe-Rn~~  &  ~~8.36~~  &  ~~8.27~~  &  ~~$-$1.08\%~~~  &  8.31  &   ~~$-$0.60\%~~~  &   8.27  &   ~~$-$1.08\%~~~  &  8.31  &   ~~$-$0.61\%~~  \\
\hline
\end{tabular}
\end{table}

\begin{table}[h!]
\scriptsize
\caption{\footnotesize Absolute (A.E. in kcal/mol) and relative (R.E.) error for the interaction energy of molecular dimers
from the S66 database~[37,38] obtained either with the old or new scaling law.}
\begin{center}
\begin{tabular}{|c||c|c|c|c|c|}
\hline
  SYSTEM NAME       & A.E. (OLD)  &   R.E. (OLD)  &  A.E. (NEW)  &  R.E. (NEW) &  $|$A.E. (OLD)$|-|$A.E. (NEW)$|$ \\
\hline
  WaterWater                &  $-0.35$      &   ~$-$7.1\%     &       $-0.30$    &    ~$-$6.1\%     &  $+$0.05     \\
  WaterMeOH                 &  $-0.26$      &   ~$-$4.6\%     &       $-0.16$    &    ~$-$2.8\%     &  $+$0.10     \\
  WaterMeNH2                &  $-0.96$      &   $-$13.7\%     &       $-0.86$    &    $-$12.3\%     &  $+$0.10     \\
  WaterPeptide              &  $-0.07$      &   ~$-$0.8\%     &       $+0.06$    &    ~$+$0.7\%     &  $+$0.00     \\
  MeOHMeOH                  &  $-0.26$      &   ~$-$4.5\%     &       $-0.18$    &    ~$-$3.2\%     &  $+$0.08     \\
  MeOHMeNH2                 &  $-1.06$      &   $-$14.0\%     &       $-0.93$    &    $-$12.2\%     &  $+$0.13     \\
  MeOHPeptide               &  $-0.33$      &   ~$-$3.9\%     &       $-0.25$    &    ~$-$3.0\%     &  $+$0.08     \\
  MeOHWater                 &  $-0.31$      &   ~$-$6.2\%     &       $-0.28$    &    ~$-$5.5\%     &  $+$0.04     \\
  MeNH2MeOH                 &  $-0.43$      &   $-$13.8\%     &       $-0.34$    &    $-$10.9\%     &  $+$0.09     \\
  MeNH2MeNH2                &  $-0.41$      &   ~$-$9.7\%     &       $-0.21$    &    ~$-$4.9\%     &  $+$0.20     \\
  MeNH2Peptide              &  $-0.09$      &   ~$-$1.7\%     &       $+0.08$    &    ~$+$1.5\%     &  $+$0.01     \\
  MeNH2Water                &  $-0.68$      &   ~$-$9.2\%     &       $-0.53$    &    ~$-$7.2\%     &  $+$0.15     \\
  PeptideMeOH               &  $-0.04$      &   ~$-$0.7\%     &       $+0.10$    &    ~$+$1.6\%     &  $-$0.06     \\
  PeptideMeNH2              &  $-0.71$      &   ~$-$9.4\%     &       $-0.48$    &    ~$-$6.4\%     &  $+$0.23     \\
  PeptidePeptide            &  $-0.24$      &   ~$-$2.7\%     &       $-0.06$    &    ~$-$0.7\%     &  $+$0.18     \\
  PeptideWater              &  $-0.03$      &   ~$-$0.7\%     &       $+0.02$    &    ~$+$0.4\%     &  $+$0.02     \\
  UracilUracilBP            &  $-0.19$      &   ~$-$1.1\%     &       $-0.12$    &    ~$-$0.7\%     &  $+$0.07     \\
  WaterPyridine             &  $-0.77$      &   $-$11.1\%     &       $-0.68$    &    ~$-$9.8\%     &  $+$0.09     \\
  MeOHPyridine              &  $-0.73$      &   ~$-$9.8\%     &       $-0.65$    &    ~$-$8.7\%     &  $+$0.08     \\
  AcOHAcOH                  &  $-0.89$      &   ~$-$4.6\%     &       $-0.82$    &    ~$-$4.2\%     &  $+$0.07     \\
  AcNH2AcNH2                &  $-0.19$      &   ~$-$1.2\%     &       $-0.09$    &    ~$-$0.5\%     &  $+$0.10     \\
  AcOHUracil                &  $-0.42$      &   ~$-$2.1\%     &       $-0.35$    &    ~$-$1.8\%     &  $+$0.07     \\
  AcNH2Uracil               &  $-0.14$      &   ~$-$0.7\%     &       $-0.05$    &    ~$-$0.2\%     &  $+$0.09     \\
  BenzeneBenzenepipi        &  $-0.71$      &   $-$25.7\%     &       $-0.64$    &    $-$23.2\%     &  $+$0.07     \\
  PyridinePyridinepipi      &  $-0.70$      &   $-$18.4\%     &       $-0.59$    &    $-$15.3\%     &  $+$0.12     \\
  UracilUracilpipi          &  $+0.02$      &   ~$+$0.2\%     &       $+0.23$    &    ~$+$2.3\%     &  $-$0.21     \\
  BenzenePyridinepipi       &  $-0.73$      &   $-$21.6\%     &       $-0.63$    &    $-$18.7\%     &  $+$0.10     \\
  BenzeneUracilpipi         &  $-0.37$      &   ~$-$6.6\%     &       $-0.19$    &    ~$-$3.4\%     &  $+$0.18     \\
  PyridineUracilpipi        &  $-0.19$      &   ~$-$2.8\%     &       $-0.02$    &    ~$-$0.2\%     &  $+$0.17     \\
  BenzeneEthene             &  $-0.58$      &   $-$41.6\%     &       $-0.52$    &    $-$37.3\%     &  $+$0.06     \\
  UracilEthene              &  $-0.31$      &   ~$-$9.2\%     &       $-0.13$    &    ~$-$3.8\%     &  $+$0.18     \\
  UracilEthyne              &  $-0.07$      &   ~$-$1.8\%     &       $+0.05$    &    ~$+$1.3\%     &  $+$0.02     \\
  PyridineEthene            &  $-0.55$      &   $-$30.2\%     &       $-0.45$    &    $-$24.6\%     &  $+$0.10     \\
  PentanePentane            &  $-1.29$      &   $-$34.5\%     &       $-0.86$    &    $-$23.0\%     &  $+$0.43     \\
  NeopentanePentane         &  $-0.76$      &   $-$29.1\%     &       $-0.54$    &    $-$20.7\%     &  $+$0.22     \\
  NeopentaneNeopentane      &  $-0.66$      &   $-$37.8\%     &       $-0.57$    &    $-$32.7\%     &  $+$0.09     \\
  CyclopentaneNeopentane    &  $-0.95$      &   $-$39.9\%     &       $-0.73$    &    $-$30.5\%     &  $+$0.22     \\
  CyclopentaneCyclopentane  &  $-1.11$      &   $-$37.1\%     &       $-0.80$    &    $-$27.0\%     &  $+$0.30     \\
  BenzeneCyclopentane       &  $-0.80$      &   $-$22.6\%     &       $-0.50$    &    $-$14.2\%     &  $+$0.30     \\
  BenzeneNeopentane         &  $-0.49$      &   $-$17.2\%     &       $-0.30$    &    $-$10.5\%     &  $+$0.19     \\
%
%
  UracilPentane             &  $-0.55$      &   $-$11.5\%     &       $-0.16$     &    ~$-$3.3\%       &  $+$0.40     \\
  UracilCyclopentane        &  $-0.55$      &   $-$13.4\%     &       $-0.23$     &    ~$-$5.7\%       &  $+$0.32     \\
  UracilNeopentane          &  $-0.41$      &   $-$11.1\%     &       $-0.15$     &    ~$-$4.2\%       &  $+$0.26     \\
  EthenePentane             &  $-0.67$      &   $-$33.5\%     &       $-0.36$     &    $-$18.3\%       &  $+$0.30     \\
  EthynePentane             &  $-0.58$      &   $-$33.5\%     &       $-0.49$     &    $-$28.4\%       &  $+$0.09     \\
  PeptidePentane            &  $-0.73$      &   $-$17.3\%     &       $-0.28$     &    ~$-$6.6\%       &  $+$0.45     \\
  BenzeneBenzeneTS          &  $-0.05$      &   ~$-$1.7\%     &       $+0.11$     &    ~$+$4.0\%       &  $-$0.07     \\
  PyridinePyridineTS        &  $+0.04$      &   ~$+$1.0\%     &       $+0.23$     &    ~$+$6.5\%       &  $-$0.19     \\
  BenzenePyridineTS         &  $-0.02$      &   ~$-$0.6\%     &       $+0.15$     &    ~$+$4.4\%       &  $-$0.12     \\
  BenzeneEthyneCHpi         &  $-0.01$      &   ~$-$0.3\%     &       $+0.11$     &    ~$+$4.0\%       &  $-$0.10     \\
  EthyneEthyneTS            &  $-0.24$      &   $-$15.5\%     &       $-0.19$     &    $-$12.6\%       &  $+$0.04     \\
  BenzeneAcOHOHpi           &  $+0.10$      &   ~$+$2.2\%     &       $+0.20$     &    ~$+$4.3\%       &  $-$0.10     \\
  BenzeneAcNH2NHpi          &  $-0.07$      &   ~$-$1.7\%     &       $+0.08$     &    ~$+$1.8\%       &  $-$0.01     \\
  BenzeneWaterOHpi          &  $-0.34$      &   $-$10.3\%     &       $-0.22$     &    ~$-$6.7\%       &  $+$0.12     \\
  BenzeneMeOHOHpi           &  $-0.40$      &   ~$-$9.6\%     &       $-0.25$     &    ~$-$6.0\%       &  $+$0.15     \\
  BenzeneMeNH2NHpi          &  $-0.33$      &   $-$10.2\%     &       $-0.18$     &    ~$-$5.5\%       &  $+$0.15     \\
  BenzenePeptideNHpi        &  $-0.17$      &   ~$-$3.3\%     &       $+0.01$     &    ~$+$0.3\%       &  $+$0.16     \\
  PyridinePyridineCHN       &  $+0.45$      &   $+$10.7\%     &       $+0.52$     &    $+$12.3\%       &  $-$0.07     \\
  EthyneWaterCHO            &  $-0.14$      &   ~$-$4.9\%     &       $-0.10$     &    ~$-$3.4\%       &  $+$0.04     \\
  EthyneAcOHOHpi            &  $-0.24$      &   ~$-$4.9\%     &       $-0.16$     &    ~$-$3.3\%       &  $+$0.08     \\
  PentaneAcOH               &  $-0.73$      &   $-$25.2\%     &       $-0.47$     &    $-$16.4\%       &  $+$0.25     \\
  PentaneAcNH2              &  $-0.72$      &   $-$20.4\%     &       $-0.44$     &    $-$12.6\%       &  $+$0.27     \\
  BenzeneAcOH               &  $-0.34$      &   ~$-$8.9\%     &       $-0.12$     &    ~$-$3.1\%       &  $+$0.22     \\
  PeptideEthene             &  $-0.37$      &   $-$12.2\%     &       $-0.13$     &    ~$-$4.3\%       &  $+$0.24     \\
  PyridineEthyne            &  $-0.36$      &   ~$-$9.0\%     &       $-0.29$     &    ~$-$7.2\%       &  $+$0.07     \\
  MeNH2Pyridine             &  $-0.28$      &   ~$-$7.0\%     &       $-0.09$     &    ~$-$2.3\%       &  $+$0.19     \\
\hline
\end{tabular}
\end{center}
\end{table}


\begin{thebibliography}{99}
\bibitem{Comment_4} The proportionality function
can be also written as $C (\mu\omega, R_{\rm vdW}) = \frac 12 \left[ \left(\frac{4}{\pi^5}\right)^{\nicefrac 13}
\frac{3}{n (R_{\rm vdW}) V^{\nicefrac 13}} \right]^{\nicefrac 27}$. Here, the volume $V = \int r^3 n (\mathbf{r})
d \mathbf{r} = (\frac{16}{\pi})^{\nicefrac 12}
(\frac{\hbar}{\mu\omega})^{\nicefrac 32}$ is occupied by the charge density
$n (\mathbf{r}) = (\frac{\mu\omega}{\pi\hbar})^{\nicefrac 32} e^{-\frac{\mu\omega}{\hbar} r^2}$
of the QDO and $n (R_{\rm vdW})$ is its value at the vdW radius.
\bibitem{Comment_5} A strict derivation of the proportionality constant for real many-electron atoms
requires additional work and it is a subject of our current studies.
\bibitem{Comment_7} The multipole polarizabilities are defined as
$\alpha_1 \equiv \alpha_{\rm dipole}$, $\alpha_2 \equiv \alpha_{\rm quadrupole}$,
$\alpha_3 \equiv \alpha_{\rm octupole}$, and so on.
\end{thebibliography}
\end{document}